\documentclass[12pt,preprint]{aastex}
\usepackage[dvips]{epsfig}
\usepackage{lscape}
\newcommand{\hii}{{\sc H~ii}}
\newcommand{\Hb}{{H$\beta$}}
\newcommand{\Fb}{$F$(H$\beta$)}
\newcommand{\Ib}{$I$(H$\beta$)}
\newcommand{\ew}{$EW$(H$\beta$)}
\newcommand{\OIII}{$I(\lambda\, 5007)/I$(H$\beta$)}
\newcommand{\OII}{$I(\lambda\, 3727)/I$(H$\beta$)}

\newcommand{\oIII}{$I(\lambda\, 5007)$}
\newcommand{\oII}{$I(\lambda\, 3727)$}

\newcommand{\sII}{$I(\lambda\, 6720)$}
\newcommand{\sIIratio}{$I(\lambda\, 6717)/I(\lambda\, 6731)$}
\newcommand{\oiii}{{\sc [O iii]}}
\newcommand{\oii}{{\sc [O ii]}}
\newcommand{\sii}{{\sc [S ii]}}

\newcommand{\cs}{$\lambda\, 5007$}
\newcommand{\tsvs}{$\lambda\, 3727$}
\newcommand{\csos}{$\lambda\, 4686$}
\newcommand{\ssv}{$\lambda\, 6720$}
\newcommand{\ctst}{$\lambda\, 4363$}
\newcommand{\stcc}{$\lambda\, 6300$}

\newcommand{\mup}{$M_{up}$}

\newcommand{\beq}{\begin{equation}}
\newcommand{\eeq}{\end{equation}}
\newcommand{\beqar}{\begin{eqnarray}}
\newcommand{\eeqar}{\end{eqnarray}}
\newcommand{\chb}{$C$(H$\beta$)}
\def\lesssim{\mathrel{\hbox{\rlap{\hbox{\lower4pt\hbox{$\sim$}}}\hbox{$<$}}}}
\def\grsim{\mathrel{\hbox{\rlap{\hbox{\lower4pt\hbox{$\sim$}}}\hbox{$>$}}}}
\newcommand{\qH}{$Q($H$^0$)}
\newcommand{\rdt}{$R_{23}$}

\shorttitle{PHOTOIONIZATION MODELING OF NGC~5461}
\shortauthors{\sc Luridiana and Peimbert}

\begin{document}
\title{THE STRUCTURE AND STAR-FORMATION HISTORY OF NGC~5461}\label{Cap_5461}
\author{V. Luridiana\altaffilmark{1} and M. Peimbert}
\affil{Instituto de Astronom\'\i a, Universidad Nacional Aut\'onoma de
M\'exico, \\Apartado Postal 70-264, 04510 M\'exico D.F., Mexico}
\email{vale,peimbert@astroscu.unam.mx}

\altaffiltext{1}{Present address: Observatoire Midi-Pyr\'en\'ees, Laboratoire d'Astrophysique,
14, Av. E.Belin, 31400 Toulouse, France}

\begin{abstract}
We compute photoionization models for the
giant extragalactic {\sc H ii} region NGC~5461,
and compare their predictions to several observational 
constraints.
Since we aim at reproducing not only the global properties
of the region, but its local structure also,
the models are constrained to reproduce the observed density
profile, and our analysis takes into consideration 
the bias introduced by the shapes and sizes
of the slits used by different observers.
We find that an asymmetric nebula with a gaussian
density distribution, powered by a young burst of 3.1 Myr, 
satisfactorily reproduces most of the constraints,
and that the star-formation efficiency inferred from the model 
agrees with current estimates.
Our results strongly depend on the assumed density law,
since constant density models overestimate
the hardness of the ionizing field, affecting the deduced
properties of the central stellar cluster.
We illustrate the features of our best model, 
and discuss the possible sources of errors and uncertainties
affecting the outcome of this type of studies.
\end{abstract}


\section{INTRODUCTION}

Tailored photoionization models have proven to be a useful tool
for the understanding of star-formation (SF) regions
(e.g., Garc\'\i a-Vargas et al. 1997; Gonz\'alez-Delgado \& P\'erez 2000;
Luridiana, Peimbert, \& Leitherer 1999; Luridiana 1999; Stasi\'nska \& Schaerer 1999).
This has been especially true since spectral energy distributions
(SEDs) for population synthesis models have become available,
providing a much better approximation to real ionizing spectra
than the na\"\i ve one-spectrum models.

Many uncertainties still affect the output of
photoionization models:
incomplete stellar tracks, unknown geometry both of the
stellar source and the ionized nebula,
processes -other than photoionization- participating
in the gas thermal balance,
to mention only a few of them. 
An excellent review on this kind of uncertainties
can be found in Stasi\'nska (2000).
An additional source of uncertainty lies in
the stochasticity of the Initial Mass Function (IMF) in real clusters,
which leads to fluctuations of the number of stars of a given mass value
around the average (analytical) value (Cervi\~no, Luridiana \& Castander 2000). 
This source of uncertainty is intrinsic to population synthesis models
and cannot be removed, and it especially affects low-mass star clusters
as it depends on statistics;
it is probably not relevant for the case
of NGC~5461, as it will be shown in Section~\ref{mass_estimates}.

In spite of these uncertainties,
the computation of tailored photoionization models can
provide many insights into the physical processes
going on in photoionized regions. 
In the present work we describe a selected photoionization model of NGC~5461,
and illustrate the properties that can be inferred for the region,
emphasizing the uncertainties involved and the problems still unsolved.
The predictions of the model are compared to a large set of observational 
constraints, 
including both global (e.g., the total emitted \Hb\ flux) and more local properties
(e.g., the line intensity ratios for different apertures).
The comparison between the predictions and the observations
is always made after correcting the model output for the size and shape
of the aperture used in each observation.
Furthermore, we did not enforce any {\it a priori} density law, 
but rather determined a gas distribution yielding self-consistent predictions
in agreement with resolved radial observations of the \sIIratio\ doublet.
This is by far the most innovative point in our modeling procedure,
and, quite certainly, one of the most important: 
in fact, in the following we will demonstrate that
our results strongly depend on the assumed density law,
and that simpler, not-tailored gas distributions, 
such as constant density models taken from large photoionization grids, 
overestimate the hardness of the radiation field, 
leading to a significant bias in the deduction of 
the properties of the ionization source.

\section{GENERAL PROPERTIES}

NGC~5461 is a giant extragalactic {\sc H~ii} region (GEHR), 
located in one arm of the spiral galaxy M~101 (NGC~5457).
NGC~5461 has been the object of several studies
(Israel, Goss, \& Allen 1975;
Sandage \& Tammann 1976;
Rayo, Peimbert, \& Torres-Peimbert 1982;
McCall, Rybski, \& Shields 1985;
Evans 1986;
Melnick et al. 1987;
Skillmann \& Israel 1988;
Torres-Peimbert, Peimbert, \& Fierro 1989;
Casta\~neda, Vilchez, \& Copetti 1992;
Rosa \& Benvenuti 1994;
Williams \& Chu 1995;
Kennicutt \& Garnett 1996;
Garnett et al. 1999).
In the modeling, we used as many data as possible
from these sources, unless relevant information 
from the reference paper was missing.
Additionally, we used high-resolution spectroscopic data
taken in June 1996 with the 2.1 m telescope at the 
Observatorio Astron\'omico Nacional de S. Pedro M\'artir
(Luridiana, Esteban, \& Peimbert, in preparation).
In the following sections we will discuss the determinations 
of relevant parameters made by different authors.

\subsection{Radius}\label{radius}

The radius of an extended object is a somewhat tricky parameter,
since it depends on the lower limit set on the observed flux,
and on the frequency considered.
Casta\~neda et al. (1992) 
reproduce the H$\alpha$ brightness profile of NGC 5461 
with the superposition of two gaussian density distributions,
of characteristic radius $r_0=1.3''$, 
separated by $2''$ in the plane of the sky. 
Williams \& Chu (1995) show
an uncalibrated H$\alpha$ image, 
in which the maximum emission comes from an elongated region of about $30''\times 15''$.
Kennicutt \& Garnett (1996), based on data by {Scowen}, 
report a diameter of $30''$ in H$\alpha$.
Israel et al. (1975) estimate a size
of $66''\times 25''$ in H$\alpha$, 
and find that the radio emission follows the same distribution;
they state that more than 50$\%$ of the flux 
is emitted in a bulge with a FWHM of $5''$.

We took into account these data to have a gross idea 
of the radius of the region, which we estimate to be about $14''$.
At a distance of $7.4$ Mpc (Sandage \& Tammann 1976, with the suggested
correction by de Vaucouleurs 1978)
this corresponds to a linear radius of $500$ pc.
In all our calculations, we always ensured
that the total radius was not too
different from this reference value.

\subsection{Electron Density and Filling Factor\label{Ne_epsilon}}

Casta\~neda et al. (1992) found that a model following the density
law
$N_{\rm e}=N_a e^{-(r/r_{\rm 0})^2}+N_b e^{-((r-\delta)/r_{\rm 0})^2}$
reproduces the observed H$\alpha$ brightness and \sIIratio\ profiles.
Our models are constrained to reproduce their observed \sIIratio\ profile,
however we did not stick to the density law they proposed
for the following reasons:

\begin{enumerate}

\item{}  The density law cannot be calculated a priori knowing only the observed \sIIratio\ profile,
  but it must rather be determined self-consistently with the overall ionization
  structure.
  Elliott \& Meaburn (1973) give an approximate expression to estimate the
  [{\sc O ii}] density of a dust-free nebula:
  	\beq
	N_{\rm e}({\rm O\; II}]) = \frac{\int_V E_{3729}(v) dv}{\int_V E_{3726}(v) dv}\,,
	\eeq	
  where $V$ is the volume and $E_{3729}(v)$, $E_{3726}(v)$ represent the local emissivities
  of [{\sc O ii}] $\lambda\, 3729$ and [{\sc O ii}] $\lambda\, 3726$.
  Casta\~neda et al. (1992) generalize this expression to the [{\sc S~ii}] case;
  however, they neglect to include the {\sc S~ii} abundance in the integrations,
  which does not cancel out since it strongly varies from point to point;
  in this respect, it is sufficient to note that the presence of {\sc~O iii} in
  the spectrum implies the presence of  {\sc S~iv}, which has roughly the same
  ionization potential.
  The fact that the {\sc S ii} abundance is not spatially constant 
  is also noted by Casta\~neda et al. (1992), when they observe
  that {\sc S ii} is not the most important ionization state of sulfur
  in a GEHR, and that it is present only in the outskirts of the nebula.

  Stated otherwise, the observed  \sIIratio\ profile at a given point does not only
  depend on the physical conditions (electron density and temperature), but also on the 
  ionization conditions found at each projected point:
  e.g., the {\sc \sii} emission from a central high-density zone can be diluted off by
  the contribution of an extended low-density halo, being invisible in projection; 
  or, if the highest-density zone is highly ionized, it does not contain any
  {\sc S ii} and gives no contribution to the \sIIratio\ ratio. 

\item{} For simplicity, we chose a centrally symmetric density law. This choice
   implies either neglecting the secondary brightness knot, seen at
   $d=-2''$ in Fig. 13 of Casta\~neda et al. (1992), or adopting an approximate scheme to model the asymmetry.
   We adopted the second solution, as it will be explained in the following.
	
\end{enumerate}

Following Casta\~neda et al. (1992),
we adopted a gaussian form for the density law.
Within this choice, two alternatives are possible:
a) Either the ionizing source is located at the brightest spots,
or b) it is shifted with respect to them.
In principle, both solutions are possible,
since the emitted intensity depends both on the density and on the
number of ionizing photons striking a given point.
However, the \sIIratio\ ratio profile indicates that the source is
most probably located somewhere between the two emission maxima.
In fact, in the innermost, high-ionization zone of the \hii\ region no {\sc S~ii}
is expected (hence no \sii\ emission),
so that a local minimum in density (local maximum in \sIIratio)
should be observed.
Instead, the \sii\ ratio shows a definite absolute minimum at the brightest knot
location, and a secondary local minumum at the secondary knot. 
This probably implies that the stellar cluster is located between the two,
and has blown away and/or consumed out the neighboring gas,
leaving a density depression surrounded by the density peaks seen in projection.
A continuum image of the region, showing the position of the stars
with respect to the gas, would be of great help in confirming this point.

Based on these considerations,
we modeled the region with off-center gaussians, following the density law
       \beq 
	\begin{array}{lr}
          N_{\rm e }(r)=  &  \left\{
              \begin{array}{ll}
              f(r) &  \hspace{3truecm} {\rm for} \hspace{.3truecm}f(r)> N_e^{\rm min} \\
              N_e^{\rm min} &  \hspace{3truecm} {\rm for}\hspace{.3truecm} f(r)\le N_e^{\rm min},
              \end{array}
       \right.
              \end{array}
       \eeq       
where $f(r)= N_0  e^{-((r-\delta)/r_{\rm 0})^2}$,
$N_e^{\rm min} = 50$ cm$^{-3}$,
and $1.0''\le \delta \le 2.0''$ and $N_0$ are tuned to meet the observational
constraints.
Since the brightness knots produced in projection by a model of this kind
are identical, whereas the peak intensity of the secondary knot in NGC 5461 is about
one half that of the primary one (Fig. 13 in Casta\~neda et al. 1992),
we calculated asymmetric models, by adding
the two halves of two different models, ionized by the same source
but differing in the density distributions.
This configuration neglects the pressure gradients at the border
between the two halves and assumes that the diffuse fields are directed
outwards. 

The filling factor can be estimated by means of the relation:
\beq
    N_{\rm e}^2(rms)=\epsilon N_{\rm e}^2(FL),
\eeq
where $N_{\rm e}^2(rms)$ is the root mean square electron density, and $N_{\rm e}^2(FL)$ is the electron
density determined through a forbidden-line ratio.
With the values given by Torres-Peimbert et al. (1989) for $N_{\rm e}^2(\rm rms)$ and $N_{\rm e}^2([${\sc S ii}$])$,
one finds $\epsilon = 0.004$; using more recent atomic data for [{\sc S ii}] 
(Pradhan \& Peng 1995; Keenan et al. 1996) we find $\epsilon = 0.002$ (see Table~\ref{obs_data}).

In our numerical experiments, we found it difficult to simultaneously obtain the desired brightness and \sIIratio\ 
profiles with two different $N_{\rm 0}$ values for the two halves:
instead, we found rather good fits by adopting equal $N_{\rm 0}$ values 
and different filling factors ($\epsilon$) on the two sides.
Fig.~\ref{f_models} sketches a section of 
a model of this kind,
with the dot size proportional to the density, and the dot
spacing proportional to the filling factor.
The assumed positions of some of the slits used to observe the
region are also shown (see Sections~\ref{apertures} and~\ref{comparison}).

As for the filling factor behavior in each half, 
we ignore whether the choice of keeping it constant is physically based, 
nor whether it is the only possible solution: 
rather, we chose it as the simplest hypothesis leading to self-consistent predictions.
A thorough discussion on the physical meaning of such configuration
is out of the scopes of the present paper.

\subsection{$Q({\rm H}^{\rm 0})$\label{qH0}}

The rate of ionizing photons emitted by the ionizing stars can be 
computed by means of the expression:
\beq
{Q}({\rm H}^{\rm 0})=\frac{I^{\rm tot}({\rm H}\beta)4\pi d^2}{h\nu} \frac{\alpha_{\rm B}}{\alpha({\rm H}\beta)},
\eeq
where $I^{\rm tot}({\rm H}\beta)$ stands for the total dereddened \Hb\ intensity, measured in erg
sec$^{-1}$cm$^{-2}$, 
$d$ is the distance to the region, $\alpha_{\rm B}$ is the
total recombination coefficient to all levels but the
first for hydrogen case B, and $\alpha({\rm H}\beta)$ 
is the effective recombination coefficient for H$\beta$.
The assumptions underlying this expression are that no photons leak out
nor are absorbed by dust, and that case B holds. 
The effects of relaxing the first assumption depend on the
constraints adopted on the models (see Section~\ref{geometry}).

$Q({\rm H}^{\rm 0})$ values determined through observations are generally underestimated,
since most slits sample only a part of the nebula, 
so that the observed values of \Ib\ are smaller than $I^{\rm tot}(H\beta)$.
To overcome this uncertainty, we considered as many data as
possible, in a sequence of larger and larger slits.
The first column of Table~\ref{compiled} lists
the logarithmic values of the observed intensity of H$\beta$ before reddening correction 
reported by different authors, in units erg sec$^{-1}$cm$^{-2}$. 
The second column shows the logarithmic reddening correction \chb, 
and the third column lists the corresponding \qH, homogeneously 
calculated by us adopting $d=7.4$ Mpc and $T_{\rm e}=9,200$ K. 

\begin{deluxetable}{lcccclllc}
\tabletypesize{\footnotesize}
\tablecaption{Compiled Parameters of NGC~5461 and Slits' Features.\label{compiled}}
\rotate
\tablewidth{0pt}
\tablehead{
\multicolumn{1}{c}{}    & \multicolumn{4}{c}{Measured Quantities}     &             \multicolumn{3}{c}{Slits' Features} \\                   
\cline{2-5}                                                                              \cline{6-9}                         \\
\multicolumn{1}{l}{Reference}                & \multicolumn{1}{c}{Log \Fb} & \colhead{\chb} & \colhead{Log \qH\tablenotemark{a}} 
& \multicolumn{1}{c}{\ew} & \colhead{Shape} & \multicolumn{2}{c}{Size}& \multicolumn{1}{c}{Label}\\
\multicolumn{1}{c}{}                   & \multicolumn{1}{c}{(erg s$^{-1}$)} & \multicolumn{1}{c}{(dex)} & \multicolumn{1}{c}{(photons s$^{-1}$)}
& \multicolumn{1}{c}{(\AA $^{-1}$)} & \colhead{} & \colhead{($''$)} & \colhead{(pc)\tablenotemark{a}}&}
\startdata
Rayo et al. 1982            &   -12.52                  & 0.75                    &\nodata & 195     & Rectangular & 3.8 $\times$ 12.4                    & 137 $\times$ 446   \\
McCall et al 1985           &    \nodata                &\ (0.60)\tablenotemark{b}&\nodata & 249     & Rectangular & \multicolumn{2}{c}{Several slits}                         \\
Melnick et al. 1987         &\ (-11.99)\tablenotemark{c}& 0.30                    & 52.44  & 223     & Circular    & \multicolumn{2}{c}{Unknown\tablenotemark{d }}             \\
Skillmann \& Israel 1988    &   -12.16                  & 0.57                    & 52.54  & \nodata & Circular    &  \diameter\  20                      & \diameter\  720    &E\\
Torres-Peimbert et al. 1989 &   -12.46                  & 0.60                    & 52.27  & 195     & Rectangular & 3.8 $\times$ 12.4                    & 137 $\times$ 446   &B\\
Kennicutt \& Garnett 1996   &   \nodata                 & 0.36                    &\nodata & 239     & Circular    & (\diameter\  10.3?)\tablenotemark{e} & (\diameter\  371?) \\ 
Luridiana et al. in preparation\tablenotemark{f} & -12.39       & 0.43                    & 52.18  & 157     & Rectangular & 4.0 $\times$ 13.3                    & 144 $\times$ 479   &C\\
Luridiana et al. in preparation\tablenotemark{g} & \nodata      & \nodata                 &\nodata &\nodata  & Rectangular & 4.0 $\times$ 39.9                    & 144 $\times$ 1436  &D \\
Garnett et al. 1999         &   \nodata                 &\ (0.57)\tablenotemark{b}&\nodata & \nodata & Rectangular & 0.86  $\times$ 0.86                  &  31  $\times$ 31 &A
\enddata
\tablenotetext{a}{Calculated for an assumed distance of 7.4 Mpc.}
\tablenotetext{b}{Derived from the published $F({\rm H}\alpha)/F({\rm H}\beta)$, adopting the extinction law by Whitford (1958).}
\tablenotetext{c}{Derived from the published ${\cal F}({\rm H}\beta)=4 \pi d^2 F({\rm H}\beta)$, and $d=6.9$ Mpc. }
\tablenotetext{d}{Greater than the halo diameter, but not specified by the authors.}
\tablenotetext{e}{Not clearly specified in the paper.}
\tablenotetext{f}{Blue side of the spectrum (3300-6600 \AA).}
\tablenotetext{g}{Red side of the spectrum (6300-9100 \AA).}
\end{deluxetable}

We adopted for our models the maximum observed value Log $Q($H$^{\rm 0})=52.54$.
Incidentally, this table illustrates the risks of modeling a region using only one set
of observational constraints:
since the local ionization degree directly depends on the rate of ionizing photons
emitted,
models in which the number of ionizing photons emitted is substantially underestimated
cannot be realistic.

\subsection{Observed Equivalent Width of \Hb}\label{ewhb}

The fourth column of Table~\ref{compiled} lists some of the values of the equivalent width of \Hb\
(\ew) found in the literature.
Though \ew\ is a relevant quantity, being at the same time 
significative and easy to measure, 
some caution should be taken before relating
it to the age of the source.
First of all, there could be a continuum emission contribution from older stars,
not related to the present burst of star formation, lowering \ew\ 
and thus mimicking an older age for the burst.
Second, the equivalent width is, ideally, a global parameter,
while the observed value is local since the slit samples
only a part of the nebula.
If the stars are concentrated in the center of
the region, we should expect that the bigger the slit, 
the higher the observed \ew.
This trend is just the opposite of what is found in Table~\ref{compiled},
a not unusual finding in the study of GEHRs (see, e.g. the case of NGC~2363:
Gonz\'alez-Delgado et al. 1994),
probably implying continuum contamination by sparse low-mass stars.
A third problem is related to the geometry of the source.
For a given stellar population,
the observed \ew\ value is smaller for density bounded
(as compared to radiation bounded) nebulae,
since the \Ib\ value superposed to a given continuum is lower in the former case
(this is probably one of the reasons underlying the apparent lack of very young starburst 
with \ew\ values around 1000).

With these uncertainties in mind, 
we can assume the maximum observed value,
\ew=249 \AA$^{-1}$, as a representative value
for the stellar population responsible of the ionization,
corresponding to ages of order 2.5 Myr $\lesssim$ t $\lesssim$ 3.5 Myr
for the instantaneous burst SF (IBSF) law.
This age range agrees with the values determined by Rosa \& Benvenuti (1994)
adopting a Miller-Scalo IMF. 
In the continuous SF (CSF) case, the mentioned \ew\ value would
correspond to the age range 4.5 Myr $\lesssim$ t $\lesssim$ 5.5 Myr.

\subsection{Apertures}\label{apertures}

The sixth to eighth columns of Table~\ref{compiled} list the features
of the slit apertures used by each observer,
with the linear dimensions homogeneously calculated
for an assumed distance of $7.4$ Mpc to the parent galaxy.
The last column lists the labels used throughout the paper
to refer to some of these slits.

\subsection{Chemical Abundances}

The adopted chemical abundances are listed in 
Table~\ref{chem_abun}.
The abundance values for He, O, N, Ne, S and Ar used
are those derived  by Torres-Peimbert et al. (1989)
through the empirical electron-temperature based
method,
since the previously published data
(Rayo et al. 1982; McCall et al. 1985)
were not corrected for the non-linearity of the detectors
(see  Peimbert \& Torres-Peimbert 1987).

\begin{deluxetable}{lc}
\tablecaption{Chemical abundances for the $t^2=0$ case.\label{chem_abun}}
\tablewidth{0pt}
\tablehead{
\colhead{Element}    & \colhead{Abundance\tablenotemark{a}}}
\startdata
N(He)/N(H)     & 0.0904 \\
$[$O/H$]$      & --3.61 \\
$[$C/H$]$      & --3.78 \\
$[$N/H$]$      & --4.74 \\
$[$Ne/H$]$     & --4.29 \\
$[$S/H$]$      & --5.30 \\
$[$Ar/H$]$     & --5.82 \\
$Z_{\rm gas}$  & 0.0066 \\
\enddata
\footnotesize{\tablenotetext{a}{$[X_i/{\rm H}]$ = Log $N(X_i)$/$N({\rm H})$; Z$_{gas}$ is given by weight.}}
\end{deluxetable}

The C abundance is calculated  
averaging the two C/O ratios 
(-0.03 and -0.37) by Garnett et al. (1999), 
corresponding to the cases $R_{\rm v}\equiv A_{\rm v}/E(B-V)=3.1$ and $5.0$, 
and assuming log O/H ~=~--3.61 
as in Torres-Peimbert et al. (1989).

The total heavy element abundance is $Z_{\rm gas}=0.0066$ for the $t^2=0$ case, 
where $t^2$ is the temperature-fluctuation parameter (Peimbert 1967).
In our numerical experiments we also explored higher metallicity values,
but found it difficult to discriminate among them
for reasons which will be explained in the following.
Thus, the standard $t^2=0$ value was adopted in the selected model
presented in this paper.

\subsubsection{The Metallicity of NGC~5461}\label{Z_5461}

The photoionization model presented in Section~\ref{results}
has been calculated assuming $Z_{\rm gas}=0.0066$ and $Z_{*}=0.008$.
The $Z_{*}=0.008$ value assumes that approximately 20$\%$ of the heavy elements
are locked in dust in the gas (Esteban et al. 1998);
it also has the (purely technical) virtue of allowing us to avoid making
interpolations between spectra of different metallicities, 
since it is one of the five $Z$ values of the Starburst99 library. 
Nevertheless, a higher metallicity is in principle possible, given the following 
considerations:

\begin{itemize}

\item[a)]
If there are temperature fluctuations in the object, 
the metallicity is underestimated if these are not taken into account
(Peimbert 1967).

\item[b)] 
The metallicity can be estimated from the Log $R_{\rm 23}$ vs. [O/H] diagram
(Pagel et al. 1979; McGaugh 1991) when the ionization parameter $U$ is known.
With the oxygen line intensities listed in Table~\ref{obs_data}
one obtains $0.78\le{\rm Log} R_{\rm 23} \le 0.84$.
Using the relationship by D\'\i az et al. (1991) between $U$ and the intensity of the sulfur lines:
      \beq 
         {\rm Log}\, U = -1.69\; {\rm Log}\;\Biggl(\frac{I(\lambda\lambda\, 6717, 6731)}{I(\lambda\lambda\, 9069, 9532)}\Biggr) -2.99
      \eeq
one obtains $0.004 \le U \le 0.019$,
with the exact value depending on the atomic parameters 
(Pradhan \& Peng 1995; Shaw \& Dufour 1994) and the data set used.
With these $R_{\rm 23}$ and $U$ values,
the calibration by McGaugh (1991) gives $-3.4\le [$O/H$] \le -3.2$, 
corresponding to the range $0.010 \le Z_{\rm gas} \le 0.017$.

\item[c)] Our model adjusts the \rdt, implying $Z_{\rm gas} \sim 0.0066$
in apparent contradiction with McGaugh's calibration. 
However, it should be taken into account that
 NGC~5461 lies close to the bend of the $R_{\rm 23}$ diagram, 
 so that the $Z$ solutions are nearly degenerate with respect to $R_{\rm 23}$. 
 Furthermore, $U$ is a parameter which should be used with caution:   
though it is usually treated as a global property of the region, 
   it is rather a local quantity. The observed $U$ values are situated
   somewhere in between a local and a global definition, since they
   are averages over finite volumes (see also Section~\ref{results}).
 The validity of the McGaugh (1991) calibration itself is uncertain.
   One reason for this lies in the ambiguity in the definition of $U$,
   discussed in the previous point.
   A second reason is that the calibration was made from a
   given set of ab initio simple photoionization models, 
   and it is difficult to estimate {\it a priori} 
   whether such calibration can be applied to real \hii\ regions. 

\item[d)] A different metallicity calibrator could  be used to resolve the
   degeneracy in \rdt, such as the [{\sc N~ii}] vs H$\alpha$ indicator proposed by van Zee et al. (1998):
   \beq
	 12 + {\rm Log}\, {\rm (O/H)}=1.02\, {\rm Log}\, ([{\rm N~II}]\,6548, 6584/{\rm H}\alpha) + 9.36,
	\eeq	
   which yields ${\rm Log \,O/H}= 8.58$ and $8.59$ 
   (that amount to $Z_{\rm gas}$ = 0.0105) with the Torres-Peimbert et al. (1989)
   and the Luridiana et al. (in preparation) data respectively;
   this result seem to confirm the hypothesis of a higher metallicity. 

\end{itemize}

We leave the metallicity question open for the moment,
since a definite answer to it should come from different criteria
than those taken into consideration in the present study. 
We will briefly discuss the variation of our models with metallicity in Section~\ref{discussion}.

\subsection{Line Intensities}

The observed line intensities are listed in Table~\ref{obs_data},
together with other derived physical parameters.
To compare with the models' predictions, 
we used the spectroscopic data by Garnett et al. (1999),
Torres-Peimbert et al. (1989), and Luridiana et al. (in preparation).
We also used the \Ib\ value by Skillmann \& Israel (1988) to map
the nebular emission with larger apertures.

\begin{deluxetable}{r@{}lccccc}
\tablecaption{Observational data.\label{obs_data}}
\tablewidth{0pt}
\tablehead{
\colhead{} &\colhead{}                         & \multicolumn{5}{c}{Observer} \\
 \cline{3-7} \\
\multicolumn{2}{c}{Quantity}      & \colhead{Gal99\tablenotemark{a}} & \colhead{TPPF89\tablenotemark{b}} & \colhead{LEP01\tablenotemark{c}} & \colhead{LEP01\tablenotemark{c}} & \colhead{SI88\tablenotemark{d}}\\
                                & & \colhead{(Slit A)} & \colhead{(Slit B)} & \colhead{(Slit C)} & \colhead{(Slit D)}& \colhead{(Slit E)}}
\startdata
$[${\sc O~ii}$]$  &~$I(\lambda\, 3727)/I$(H$\beta$)       & 2.980  & 2.138  & 1.977  &\nodata &\nodata \\
$[${\sc S~ii}$]$  &~$I(\lambda\, 4074)/I$(H$\beta$)       &\nodata & 0.013  & 0.018  &\nodata &\nodata \\
$[${\sc O~iii}$]$ &~$I(\lambda\, 4363)/I$(H$\beta$)       &$<$0.040& 0.015  & 0.011  &\nodata &\nodata \\
  $He{\sc ~ii}$   &~$I(\lambda\, 4686)/I$(H$\beta$)       &\nodata & 0.021  &\nodata &\nodata &\nodata \\
$[${\sc O~iii}$]$ &~$I(\lambda\, 5007)/I$(H$\beta$)       & 3.110  & 3.020  & 3.112  &\nodata &\nodata \\
$[${\sc O~i}$]$   &~$I(\lambda\, 6300)/I$(H$\beta$)       &\nodata & 0.015  & 0.022  &\nodata &\nodata \\ 
$[${\sc S~ii}$]$  &~$I(\lambda\, 6720)/I$(H$\beta$)       & 0.475  & 0.225  &\nodata & 0.293  &\nodata \\ 
$[${\sc O~iii}$]$ &~$I(\lambda\, 4363)/I(\lambda\, 5007)$ &$<$0.013& 0.005  & 0.004  &\nodata &\nodata \\
$[${\sc S~ii}$]$  &~$I(\lambda\, 6717)/I(\lambda\, 6731)$ &\nodata & 1.202  &\nodata & 1.321  &\nodata \\
&Log~$Q({\rm H}^{\rm 0})$                                 &\nodata &  52.27 & 52.18  &\nodata &  52.54 \\
&EW$(H\beta)$                                             &\nodata &  195   &  157   &\nodata &\nodata \\
&Log \rdt\tablenotemark{e}                                &  0.84  &  0.79  &  0.78  &\nodata &\nodata \\
&T$_{\rm e}$([{\sc O~iii}])\tablenotemark{f}              &\nodata &  9200  &  8400  &\nodata &\nodata \\
&N$_{\rm e}$([{\sc S~ii}])\tablenotemark{f}               &\nodata &   370  &\nodata &  150   &\nodata \\
&N$_{\rm e}$([{\sc O~ii}])\tablenotemark{f}               &\nodata &\nodata &  190   &\nodata &\nodata \\
\enddata
{\footnotesize
\tablenotetext{a}{Garnett et al. 1999.}
\tablenotetext{b}{Torres-Peimbert et al. 1989.}
\tablenotetext{c}{Luridiana et al., in preparation.}
\tablenotetext{d}{Skillmann \& Israel 1988.}
\tablenotetext{e}{\rdt = ($I(\lambda\lambda\, 4959, 5007)$ + \oII)/$I({\rm H}\beta).$}
\tablenotetext{f}{Calculated by us with the observed line ratios and homogeneous atomic parameters.}
}
\end{deluxetable}

Two of the slits considered (B and C) are very similar in size. Accordingly, we expect the corresponding
data to be similar, at least as far as the slits have been placed on exactly the same position,
and within the observational errors.
Indeed, the line ratios $I(\lambda\, 5007)/I$(H$\beta$) 
and $I(\lambda\, 3727)/I$(H$\beta$) are in good agreement; this is important, since
these lines are the most intense, and they dominate the ionization structure of the nebula.
$I(\lambda\, 6300)/I$(H$\beta$) shows a difference of almost 50\%,
but this line represents a  minor stage of ionization. If this line is formed 
in filaments and condensations, as it has been suggested (see, e.g., Stasi\'nska \& Schaerer 1999), 
one possible explanation for the difference between the data is that one slit 
captured more such condensations than the other.
It should also be taken into account that the line is very weak.
$I(\lambda\, 4074)/I$(H$\beta$) and $I(\lambda\, 4363)/I$(H$\beta$), also weak lines, are discrepant, 
but they agree within the observational errors.

\section{THE MODELING PROCEDURE}\label{modeling}

\subsection{The Numerical Models}\label{models}

The photoionization models of NGC~5461
have been calculated with {\sc Cloudy 90} (version 90.05; Ferland 1996).
We refer to the original documentation for a description
of the characteristics of the code.
The ionizing sources have been taken from the data set Starburst99
(Leitherer et al. 1999),
for the standard mass-loss case.
 
\subsection{Comparison with the Observational Data}\label{comparison}

Following the procedure outlined in Luridiana et al. (1999),
we corrected the models' predictions for the slit size,
before comparing them to the observational data.
This is a necessary step, since a slit samples
only a fraction of the nebula, 
in such a way that the ionization fractions, 
and thus the line intensity ratios, 
can be dramatically different from those of the complete model.
The features of the slits used by Garnett et al. (1999) (slit A), 
Torres-Peimbert et al. (1989) (slit B),
Luridiana et al. (in preparation) (slits C and D), 
and Skillmann \& Israel (1988) (slit E), have been listed in Table~\ref{compiled}.
Given the small size of slit A,  
we expect the observational data by Garnett et al. (1999) 
to depend much more on the exact position 
and the small-scale structure of the nebula
than the other data.
Hence, in the comparison we will consider more significative
a good fit with the other data sets.

\section{THE OBSERVATIONAL CONSTRAINTS}\label{obs_constraints}

As already pointed out by Casta\~neda et al. (1992),
the modeling of a 3-D region from a 2-D image is an ill-defined
problem, since many solutions are in principle possible. 
On the other hand, a totally realistic model of a region can never be calculated, 
because taking into account all the features of the small scale 
structure of a region rapidly outpowers any computational tool, 
not to mention the fact that observational data have a finite resolution
and are affected by errors.

To find a fair middle point between the simplistic and the nihilist
points of view, one should ask himself which quantities are really relevant
and worth (as well as possible) to determine, and which are not. 
With this criteria in mind, 
we adopted the following set of observational constraints:

\begin{enumerate}
\item[a)] The relevant line intensity ratios.
\item[b)] \ew.
\item[c)] \Ib.
\item[d)] The brightness profile (cfr. Casta\~neda et al. 1992).
\item[e)] The \sIIratio\ ratio.
\item[f)] The age range inferred by \ew\ (see Section~\ref{ewhb}).
\item[g)] The degree of ionization.
\end{enumerate}

By `relevant' line ratios we mean line ratios that fulfill as many as
possible of the following requirements: {\it i)} The line is bright,
so that the observational errors do not sensibly affect the analysis;
{\it ii)} The atomic physics is well known;
{\it iii)} The line is produced by known mechanisms.
By these criteria, the \tsvs\ and the \cs\ line intensities are very important constraints,
while \ctst\ is a less robust constraint since it almost surely has a contribution from processes other
than photoionization.
The \stcc\ line is generally underpredicted in photoionization models, 
and the places and circumstances of its formation are still being understood
(see the discussion in Section~\ref{oxygen}).
The \sii\ line intensities are important mappers of the low-ionization zone,
but unfortunately the atomic physics of this ion is still not well known,
and published parameters vary a lot.	
Finally, \csos\ is in principle an important constraint since it is a tracer of massive population,
but it is subject to the many theoretical uncertainties still affecting our knowledge of W-R stars.  
\ew\ is an important constraint, because it is an age indicator,
and it is measurable with a small error;
however, its interpretation is limited by the circumstances mentioned in Section~\ref{ewhb}.
\Ib\ is directly related to the ionizing power of the source, thus it is a
constraint of primary importance.
The brightness profile is a convolution of the ionizing radiation emitted
and the gas distribution, so it is a basic constraint in any attempt to
model the three-dimensional structure of a region.
The \sii\ ratio is an important constraint, since it allows us to map
the projected density of the region; the density dependence has to be
deconvolved from the ionization-degree effects, giving clues to the
low-energy range of the spectrum.

Items a), b), c), d) and e) are quite strong constraints, since we tried to 
fulfill them  as a function of the slit shape and size,
so that each item actually consists of several interrelated constraints.
Items c) and d) refer essentially to the same quantity, 
but averaged and displayed in different manners: in the first case,
we aim at reproducing the observed fraction of \Ib\ intersected by each slit,
while in the second case we calculate the $I({\rm H}\alpha)$ profile along a nebular diameter
and compare it with the data by Casta\~neda et al. (1992).

\section{RESULTS}\label{results}

\subsection{Best-fit Model Features\label{model_features}}

Our best model has been calculated assuming a burst SF law of age 3.1 Myr,
$M_{\rm up}=80$ M$_\odot$, and a Salpeter's IMF slope $1+x=2.35$.
The rate of ionizing photons emitted, \qH, was set at the value
\qH$ = 3.47 \times 10^{52}$ photon s$^{-1}$,  as stated in Section~\ref{qH0}.
The stellar metallicity has been set to $Z_*=0.008$.

The density law is given by $N_{\rm e}=500\; e^{-((r-\delta)/r_{\rm 0})^2}$ cm$^{-3}$
in both halves, with $r$ in parsecs, and $\delta = r_{\rm 0} = 54$ pc
(corresponding to $1.5''$ at the assumed distance of $7.4$ Mpc).
The filling factor has been set to $\epsilon=0.002$ and $\epsilon=0.005$ respectively,
to reproduce the low and high brightness peaks seen in the H$\alpha$ profile
(Casta\~neda et al. 1992).
We stress again that a lower observed \sIIratio\ ratio does not necessarily 
imply a higher density, since the observed value depends on
the contributions of all the gas parcels intercepted by a given line of sight, 
with weights depending on the local ionization and temperature conditions.
The covering factor of the model is $cf=1$.
A central hole, of radius $r_{\rm in}=20$ pc and $r_{\rm in}=35$ pc
in the low- and high-$\epsilon$ halves, is also present; 
the hole has only a minor effect on the line ratios, 
but it improves the fitting of the brightness profile between the peaks
(see also the discussion in Section~\ref{Ne_epsilon}).
The total radius takes different values in the two halves,
averaging 460 pc.
The gas metallicity of the model has been set to $Z_{\rm gas}=0.0066$.
Table~\ref{bestfit} and Fig.~\ref{f_structure}
show the features of this model.

\begin{deluxetable}{r@{}lcccccc}
\tabletypesize{\footnotesize}
\tablecaption{Best-fit Model Predictions.\label{bestfit}}
\tablewidth{0pt}
\tablehead{

\multicolumn{2}{c}{Constraint} & \multicolumn{1}{c}{Complete Model} & \colhead{Slit A\tablenotemark{a}} & \colhead{Slit B\tablenotemark{b}} & \colhead{Slit C\tablenotemark{c}} & \colhead{Slit D\tablenotemark{c}} & \colhead{Slit E\tablenotemark{d}}}
\startdata
    {\rm Log}& ~$I(\lambda\, 4861)$                   & 40.233 &     38.496   &     39.916   &     39.935   &     39.968   &     40.162   \\
$[${\sc O~ii}$]$  & ~$I(\lambda\, 3727)/I({\rm H}\beta)$ &  3.679 &      1.645   &      2.257   &      2.312   & \nodata      &    \nodata   \\
                  &                                       &        & {\bf(0.55)}  & {\bf(1.06)}  & {\bf(1.17)}  & \nodata      &    \nodata   \\
$[${\sc S~ii}$]$  & ~$I(\lambda\, 4074)/I({\rm H}\beta)$ &  0.032 &    \nodata   &      0.011   &      0.011   & \nodata      &    \nodata   \\
                  &                                       &        &    \nodata   & {\bf(0.91)}  & {\bf(0.64)}  & \nodata      &    \nodata   \\
$[${\sc O~iii}$]$ & ~$I(\lambda\, 4363)/I({\rm H}\beta)$ &  0.007 &      0.015   &      0.012   &      0.012   & \nodata      &    \nodata   \\
                  &                                       &        &{\bf($>$0.37)}& {\bf(0.79)}  & {\bf(1.05)}  & \nodata      &    \nodata   \\
$[${\rm He}~{\sc ii}$]$ & ~$I(\lambda\, 4686)/I({\rm H}\beta)$ &  0.000 &    \nodata   &      0.000   &    \nodata   & \nodata      &    \nodata   \\
                  &                                       &        &    \nodata   & {\bf(0.00)} &    \nodata   & \nodata      &    \nodata   \\
$[${\sc O~iii}$]$ & ~$I(\lambda\, 5007)/I({\rm H}\beta)$ &  1.553 &      3.195   &      2.575   &      2.529   & \nodata      &    \nodata   \\
                  &                                       &        & {\bf(1.03)}  & {\bf(0.85)}  & {\bf(0.81)}  & \nodata      &    \nodata   \\
$[${\sc O~i}$]$   & ~$I(\lambda\, 6300)/I({\rm H}\beta)$ &  0.023 &    \nodata   &      0.003   &      0.003   & \nodata      &    \nodata   \\
                  &                                       &        &    \nodata   & {\bf(0.21)}  & {\bf(0.15)}  & \nodata      &    \nodata   \\
$[${\sc S~ii}$]$  & ~$I(\lambda\, 6720)/I({\rm H}\beta)$ &  0.449 &      0.093   &      0.145   & \nodata      &  0.209       &    \nodata   \\
                  &                                       &        & {\bf(0.20)}  & {\bf(0.65)}  & \nodata      & {\bf(0.71)}  &    \nodata   \\
$[${\sc O~iii}$]$ & ~$I(\lambda\, 4363)/I(\lambda\, 5007)$ &  0.005 &      0.005   &      0.005   &      0.005   & \nodata      &    \nodata   \\
                  &                                       &        &{\bf($>$0.38)}& {\bf(1.00)}  & {\bf(1.20)}  & \nodata      &    \nodata   \\    
$[${\sc S~ii}$]$  & ~$I(\lambda\, 6717)/I(\lambda\, 6731)$ &  1.350 &    \nodata   &      1.228   &    \nodata   &  1.280       &    \nodata   \\
                  &                                       &        &    \nodata   & {\bf(1.02)}  &    \nodata   & {\bf(0.97)}  &    \nodata   \\    
\tableline
&$F^{\rm slit}/F^{\rm tot}$        &           &     \nodata  &      0.48    &      0.50    &   \nodata&     0.85     \\
&                                  &           &     \nodata  & {\bf(1.13) } & {\bf(1.14) } &   \nodata& {\bf(0.85 )} \\
&\ew                               &   263     &     \nodata  &       126    &       132    &   \nodata&   \nodata    \\
&                                  &{\bf(1.06)}&  \nodata  &  {\bf(0.65)} & {\bf(0.84)}  &   \nodata&   \nodata    \\
&\rdt                              &           &      0.77    &      0.75    &      0.75    &   \nodata&   \nodata    \\
                  &                &           & {\bf(0.92 )} &  {\bf(0.95)} & {\bf(0.96)}  &   \nodata&   \nodata    \\ 
\enddata
{\footnotesize
\tablenotetext{a}{Same aperture as used by Garnett et al. 1999.}
\tablenotetext{b}{Same aperture as used by Torres-Peimbert et al. 1989.}
\tablenotetext{c}{Same aperture as used by Luridiana et al., in preparation.}
\tablenotetext{d}{Same aperture as used by Skillmann \& Israel 1988.}
}
\end{deluxetable}

The first line lists the logarithm of the total H$\beta$ flux, in erg s$^{-1}$, emitted by the complete model,
and by the fractional volumes intercepted by the four slits.
In the following lines we list the predicted line ratios, \Ib, \ew, and \rdt, 
once again for the complete model and the four slit-biased cases.
For each constraint, the boldface values in parentheses are the ratios between the computed and the observed values:
i.e., a `perfect' model should rate pure 1's.

\subsubsection{Oxygen Lines}\label{oxygen} 
The agreement is very good for \cs\ and \tsvs, the most important lines
according to the criteria set in Section~\ref{obs_constraints}.
\stcc\ is too weak in our model, a not unusual fact in the history of photoionization
modeling, for which many possible explications have been invoked (e.g., Garc\'\i a-Vargas et al. 1997; 
Martin 1997; Stasi\'nska \& Leitherer 1996; Stasi\'nska \& Schaerer 1999). 

\subsubsection{{\sc [S ii]} \ssv}
\ssv\ is predicted too weak by about a factor of about 2:
more generally, in all our modeling attempts following an IBSF,
we systematically found \sII/\oII\ $ \sim 0.5 $(\sII/\oII)$^{\rm obs}$,
whereas CSF models of ages $\sim 5$ Myr give \sII/\oII\ $ \sim $ (\sII/\oII)$^{\rm obs}$,
indicating that the IB spectra lack flux in the range $0.7\; {\rm Ryd} \le\nu \le 1.0\; {\rm Ryd}$,
range responsible for the production of {\sc S~ii} but not of {\sc O~ii}.
The most probable explication for this result is the presence of an older and cooler
stellar population in NGC~5461, not taken into account by the extreme IB scenario.

\subsubsection{He {\sc ii} \csos}
\csos\ is completely missing in our model, which has an age just prior to the W-R phase onset.
The reason behind such a choice for the age is that the appearance of W-R stars
in the IB scenario yields a sudden hardening of the spectrum, dramatically rising the ionization degree;
furthermore, with any reasonable choice of the population parameters, 
even the \csos\ intensity rises too much. 
These circumstances might indicate that a strictly analytical treatment
of the stellar population evolution is not realistic,
and that the \csos\ flux possibly comes from only one star,
maybe slightly older than the average population. 
It should be also considered that the observed
\csos\ flux also contains a stellar contribution:
accounting for the stellar contribution,
the observed nebular value is reduced by a factor of two.
In the case of NGC~2363 (Luridiana et al. 1999), 
this effect was not taken into account
because the stellar contribution is not important 
for the age and metallicity of that region.
Summarizing, given the numerous uncertainties still affecting the theoretical modeling
of W-R and W-R$-$like stars,
the weakness of the observed \csos\ (see Table~\ref{obs_data})
and the additional uncertainties deriving from the statistical fluctuations
expected in the high-mass tail of the mass distribution
(Cervi\~no et al. 2000)
\csos\ should not be considered a robust constraint
for this object.

\subsubsection{{\sc [O~iii]} \ctst} 
The intensity of \ctst\ with respect
to \Hb\ is slightly underpredicted by our model,
with the exception of slit C
(which carries the greatest observational error).
The observational error on this line is 0.04 dex 
in the Torres-Peimbert et al. (1989) value, 
and twice that on the Luridiana et al. (in preparation) one, 
so that the two measurements agree within $1 \sigma$.
The average between the two observational values, taking the errors into account, 
would give $I(\lambda\, 4363)/I(\lambda\, 4861) \sim 0.135$,
with a model/observation ratio of $0.12/0.135\sim 0.88$.
However, if we consider the \ctst/\cs\ ratio instead, 
we find that it is fairly well reproduced by the model.

\subsubsection{{\sc [S~ii]} $\lambda\, 4074$}
Finally, the weak \sii\ $\lambda\, 4074$ line is underpredicted
by our model, but better reproduced than \ssv.

\subsubsection{Flux Fractions}
The line marked with $I({\rm H}\beta)^{\rm slit}/I({\rm H}\beta)^{\rm tot}$ is calculated from the first line, 
and shows the fraction of the H$\beta$ flux intercepted by each slit, 
as compared to the total H$\beta$ flux emitted by the complete model. 
The values in parenthesis are the corresponding observational values,
calculated as \qH/\qH$^{\rm max}$ = \qH/\qH$^{\rm Slit\, D}$;
the agreement is very satisfactory.

\subsubsection{$EW({\rm H}\beta$)}
The next line lists the predicted \ew\ values. Again, the agreement is rather satisfactory,
especially taking into account the uncertainties accompanying this quantity 
(see Section~\ref{ewhb}).

\subsubsection{\rdt}
Finally, the last line lists the calculated \rdt\ values, 
which fit the observed values very well.
\rdt\ turns out to be roughly constant for the different apertures,
in agreement with the result found by Kennicutt \& Garnett (1996) 
on the observational side.
However, we must caution that this is true only as far as the observed volumes
simultaneously sample low- and high-ionization zones, since the {\it local}
\rdt\ value spans along the nebula a range of more than one order of magnitude,
reflecting the large variations of the local ionization parameter.
Further resolved spectroscopic studies are needed to assess the line-of-sight
variations in a real nebula. 
For the moment, it is safe to say that \rdt\ should not be considered strictly constant,
or, equivalently, that $U$ is a local parameter.
A commonly used definition of $U$, based on the inner conditions of the nebula, 
(i.e., $U={Q(H^{\rm 0})}/{4\pi c R_{\rm in}^2 N_{\rm e}}$), is not very useful,
since it does not take into account the density distribution.
As an extreme (but not unrealistic) example, when we tune $R_{\rm in}$ from, say,
0.1 pc to 1 pc, the model stays essentially the same, while $U$ varies
by 2 orders of magnitude.

\subsubsection{Brightness and {\sc [S ii]} emission profiles}
In Fig.~\ref{f_profile} the predicted H$\alpha$ and \sIIratio\ ratio profiles (solid lines)
are compared to their observational counterparts.
The zero-point of the computed profiles has been shifted to $r=-1''$ (36 pc) to make it 
coincide with the zero-point set for the observational data in the original paper.

The brightness profile was calculated assuming a slit width of 36 pc ($1''$).
This is presumably the aperture used by Casta\~neda et al. (1992), 
as obtained by comparing their instrumental FWHM intensity (1.88 \AA) with
the sulfur lines FWHM intensity ($\sim$ 2.8 \AA), 
and taking into account the dispersion (0.71 \AA\ pixel$^{-1}$) and the spatial scale
(0.33 $''$ pixel$^{-1}$) reported in the original paper.
Fortunately, the profile is quite insensitive to the exact slit value used, 
at least for aperture values smaller than $3''$.
We intentionally did not smooth the calculated profile, to show that the model 
faithfully reproduces the observed data in a very local sense;
nevertheless, averaging the intensities values over $1''$ intervals,
following the observational procedure, would undoubtedly improve the fitting.

The agreement of our model with the observed \sIIratio\ values is also very good.
The computed profile clearly shows how two exactly identical local density distributions
can give rise to different \sIIratio\ ratios, only by virtue of different $\epsilon$ values.
We also note that the observed \sIIratio\ value at $d=+4''$ is clearly unphysical.

\subsection{Total stellar and ionized gas mass estimates\label{mass_estimates}}

The model described in Section~\ref{model_features} yields
a stellar mass of $M_*^{M>0.8{\rm M}_\odot} = 1.0 \times 10^{6}$ M$_\odot$ between 0.8 and 80 M$_\odot$,
or, equivalently, $M_*^{M>1.0{\rm M}_\odot} = 0.9 \times 10^{6}$ M$_\odot$ between 1.0 and 80 M$_\odot$.
Rosa \& Benvenuti (1994),
observing NGC~5461 with a $1''$ aperture,
fitted the observed continuum to that of population synthesis models,
after correcting for the attenuation by dust and the nebular continuum emission. 
Following this method, they estimated a stellar mass of 
$M_*^{M>2.0{\rm M}_\odot}(1'') = 1.0 \times 10^{5}$ M$_\odot$ in the $(2.0-80)$ M$_\odot$ range,
corresponding to $M_*^{M>1.0{\rm M}_\odot}(1'') = 1.4 \times 10^{5}$ M$_\odot$
with a Salpeter IMF slope.
Adopting a scale factor of 4 to account for the small aperture used
(Giannakopoulou-Chreighton, Fich, \& Wilson 1999),
this figure translates into  $M_*^{M>1.0{\rm M}_\odot} = 5.6 \times 10^{5}$ M$_\odot$,
i.e. roughly 2/3 of our estimate.

Carigi, Col\'\i n, \& Peimbert (1999) propose IMFs, based on the one
by Kroupa, Trout, \& Gilmore (1993),
accounting for dark matter in the form of substellar bodies.
Their IMFs are parameterized as a function of an 
$r$-value\footnote{This $r$ has no relation with the $r$ defined in Section~\ref{radius}}, 
which depends on the assumed slope in the $M < 0.5$ M$_\odot$ range.
Their preferred IMF (corresponding to the $r=1.8$ case: see their paper),
truncated at $M_{up} = 80 $ M$_\odot$, yields:
\beq
      \frac{M_{1.0}^{80}}{M_{0.01}^{80}}= 0.206,
\eeq
while with the IMF recently determined by Kroupa (2000)
one obtains for the same mass range:
\beq
      \frac{M_{1.0}^{80}}{M_{0.01}^{80}}= 0.377. 
\eeq 
These relations allow to estimate the total stellar mass of our model region: 
\beq
M_*^{tot} \simeq M_*^{M>1.0{\rm M}_\odot} / 0.292 = 3.08 \times 10^{6}{\rm M}_\odot 
\eeq

The total ionized mass obtained through straight integration of the local 
density over the volume is  
\beq
M_{\rm gas}^{\rm tot} = 1.63 \times 10^{6}{\rm M}_\odot.
\eeq
Observationally, this expression corresponds to the mass estimated
through forbidden-line density:
\beqar
M_{\rm gas}^{\rm FL} & = & A \int_V \epsilon N_e(FL)dv \\
                 & = & A \int_V \epsilon^{1/2} N_e(rms)dv, \label{M_FL}
\eeqar
where 
$A=m_H \Bigl[{4N({\rm He})}/{N({\rm H})+1}\Bigr] \Bigl[{N({\rm H})}/\Bigl({N({\rm H})+N({\rm He})\Bigr)}\Bigr]$.
A different observational estimate of the total mass of the ionized gas can be made
by integrating the rms electronic density over the total volume:
\beq
M_{\rm gas}^{\rm rms}= A \int_V N_e(rms)dV \label{M_rms}
\eeq
Peimbert (1966) showed that the expressions~\ref{M_FL} and~\ref{M_rms}
represent lower and higher limits to the total ionized mass value,
since in real nebulae the density contrast is not as extreme as supposed
by the filling-factor scheme.
In our case, this implies for the total mass of the ionized gas in NGC~5461:
\beq
1.63 \times 10^{6}{\rm M}_\odot<M_{\rm gas}^{\rm NGC~5461}< 3.25 \times 10^7 {\rm M}_\odot.
\eeq
 
According to recent estimations by Giannakopoulou-Chreighton et al. (1999),
the total molecular mass in NGC~5461 lies in the $(15-40) \times 10^6$ M$_\odot$ range,
accompanied by 1-2 times as much neutral mass.
Taking an average value of $6 \times 10^7$  M$_\odot$
for the sum of these two components,
and adding the ionized gas value, 
we find a total gaseous mass in the $(6-9) \times 10^7$  M$_\odot$ range.
The ratio of the total stellar mass to the total gaseous mass
yields a star-formation efficiency in the (0.03-0.05) range,
in agreement with current estimates of this parameter
(e.g., Lada 1992; Evans \& Lada 1991).

We conclude by remarking that the rather high value found for $M_*^{tot}$
imply that statistical fluctuations of the IMF
do not play a significant role in this region
(see also Cervi\~no et al. 2000).

\section{DISCUSSION}\label{discussion}

In this Section, we wish to justify our choice for the parameters of our favoured model, 
by means of schematically illustrating the changes in the results
obtained through variations in the input ingredients:

\subsection{Age}

The constraints on the age are set by the observed \ew\ value,
and by the ionization degree of the nebula. 
For an IB scenario, the \oIII/\oII\ ratio steadily decreases with age
until $t\sim 3$ Myr, with the exact age value depending on metallicity
and on the other stellar population parameters. 
Then, the W-R stars are born and abruptly increase the ionization degree,
which falls again at about 5 Myr following the death of W-R stars.
Thus, in the $0-5$ Myr window there is only a short period around $3$ Myr
compatible with the observations.
Ages greater than 5 Myr are excluded by the high \ew\ observed.

The CSF case, which allows older ages, will be discussed in Section~\ref{SFR}.

\subsection{Metallicity}

The results presented here are scarcely dependent on metallicity,
since at $Z\sim $ Z$_{\odot}/3$ the increase in the number
of emitters is almost perfectly counterbalanced by the decrease in 
electron temperature, leading to almost constant oxygen line ratios.
Different metallicities require slightly different age values, 
mainly because the evolution of the ionizing flux depends on mass-loss
rates, which in turn depends on $Z$.

\subsection{Geometry}\label{geometry}

The chosen geometry was the result of many crossed observational constraints:
mainly the \sIIratio\ profile, $\epsilon$, the ionization degree, the nebula 
radius, and the brightness profile.
Although we cannot ensure that the solution is unique, we are confident
that at least qualitatively the real gas distribution is not too far
from our model's assumptions. 
Nevertheless, it is useful to discuss possible variations in the input
parameters determining geometry.

\subsubsection{Density}
A variation in the density normalization ($N_0$), 
leaving the remaining parameters unchanged, 
yields modifications in both the total radius
and the ionization degree of the nebula.
The total radius changes due to the constraint on \qH, 
the ionization degree changes due to the variations in the
ionization and recombination rates.
The recombination rate increases roughly with the squared density,
while the ionization rate increases only linearly with the density,
implying that a rise in $N_0$ at fixed radius yields a fall in the ionization degree.
On the other hand, the ionization rate depends on the rate of 
photons striking a given point, which in turns depends 
also on the distance from the source; 
the average distance is smaller
in a higher-density model if \qH\ is fixed.
In our configuration, the density effect outweighs
the distance effect, so that an increase in $N_0$ lowers the ionization degree.

An interesting question that can be raised on this subject is
how the model would change, had a simpler density structure been adopted
(e.g., a constant density or hollow sphere, such as those available from
 grids of photoionization models).
The answer is that the features of our best model are strongly dependent
on the density structure chosen. 
Given the size of the region, a constant density model must be characterized
by a very low density,
with the two opposite effects mentioned in the previous discussion
(fall in the recombination rate due to the lower average density,
fall in the ionization rate due to the higher average distance).
As a rule, for a given ionization source, the average ionization degree drops 
in constant density models. 
A second major change, with respect to our gaussian density-distribution models,
is that the brightness profile is no longer reproduced.
Several constant-density models were calculated to confirm this predictions.

The implication is that, dropping both constraints on the region's radial structure
(the \sIIratio\ and brightness profiles),
a harder ionization source would be invoked to reproduce the observed line intensities,
leading to a strong bias in the inference of the central star cluster properties.
Summarizing, we claim that {\it to assess the properties of the ionizing field it is necessary
to use tailored density distributions}; this is perhaps the more important
result of the present study.

\subsubsection{Filling Factor}
The filling factor acts on the ionization degree via the average distance
from the ionizing stars.
Since the recombination rate depends on the local (clump) density only,
an increase in $\epsilon$, with the other parameters kept constant,
lowering the average distance of the gas parcels from the ionizing source,
yields a higher ionization degree.
(Note that this is true only as far as the nebula is radiation bounded: 
the inverse trend can be found in density bounded objects with fixed radius,
in which the higher the concentration towards the center, 
the less the low-ionization zone is `sacrificed' by the constraint on the radius.)

\subsection{Density-bounded Models}

The density-bounded case can be described through the two 
limiting cases of a covering factor smaller than 1,
and of a spherically symmetric nebula with $R<R_S$, where $R_S$ is the Str\"omgren radius.

In the first case the gas covers a solid angle $\Omega<4\pi$
(the covering factor is defined as $cf=\Omega/4\pi$),
i.e. the nebula is not spherically symmetric;
the ionizing photons emitted in some directions are completely absorbed,
while those emitted in other directions escape, resulting 
in a photon leakage independent of frequency.

In the second case, the nebula is spherically symmetric, 
and the photon leakage affects preferentially the highest frequencies,
i.e. those frequency characterizing the photons reaching further into the gas.
The common features between these two situations are the following:
a) the \ew\ is lower with respect to a radiation-bounded case, and
b) if \Fb\ is fixed as a constraint, \qH\ must be increased in density-bounded
models with respect to ionization-bounded models.
In the following, the two cases will be discussed separately;
of course, real nebulae are intermediate between them.

\subsubsection{Covering Factor}
In the models with $cf<1$,
we scaled the rate of ionizing photons according to the
relationship $Q'($H$^{\rm 0})=Q^0($H$^{\rm 0})/cf$,
in order to preserve the total emitted \Hb\ flux.
The consequences of assuming $cf<1$ in the models depend on which other parameters, 
if any, are correspondingly modified. 
This, in turns, depends on the obsevational constraints set.
One of the most stringent constraints of the present work
is the observed \Ib\ as a function of the aperture used.
Starting from our reference model with $cf=1$, any decrease in $cf$ should be
associated to a corresponding increase in \qH\ to preserve the \Hb\ flux
seen through each slit. 
This would lead to an increase in the ionization degree, calling
for some further change in the stellar source and/or in the gas
geometry to be balanced. 
We found that moderate changes in $cf$ (say, lowering $cf$ from $1$ to $0.75$)
do not alter sensibly the important line ratios. 
The only changes are the following:
\ew\ decreases, getting farther from the observed values;
\Ib\ slightly decreases, but the effect can be easily counterbalanced
from a small modification in $N_{\rm e}^{min}$;
and {$I(\lambda\, 4074)$/$I(\lambda\, 4861)$} improves
due to the slight increase in temperature.
On the other hand,
{$I(\lambda\, 6720)$}
does not improve since the the ionization structure is essentially 
the same, and  this line is less sensitive then {$\lambda\, 4074$} to changes in temperature.

\subsubsection{External Radius}

If the external radius is truncated before $R_S$ is reached, 
the overall ionization degree of the nebula is altered,
since the low-ionization lines are formed in the outskirts of the nebula.
A detailed discussion on these models is quite complex, due to the 
interrelation of all the constraints, which implies that a change 
in one input parameter calls for changes in other parameters as well.
Necessarily, our discussion will be simplified in this respect,
and we will try to illustrate separately the cascade of consequences
which results from relaxing the assumption $R=R_S$.

The simplest possible solution is to simply truncate the nebula
before the Str\"omgren radius, without further changes. 
The resulting models will generally preserve the value of most considered line ratios,
namely \oiii, \oii, \sii, and He~{\sc ii} lines. 
This is a consequence of the ionization structure of the nebula,
of the density distribution, 
and of the size of the slits as compared to the size of the region:
these three factors contribute in the same direction
(we are not considering here extreme cases of very small radii).
The only line which is strongly affected is \stcc, which was already
underpredicted. 
A major problem with models of this kind is that they do not reproduce the constraints
on the observed brightness profile, which becomes too weak, 
and on the radius, which becomes too small.
The constraint on the radius may then be fulfilled by a decrease in density and/or
filling factor, but the problem with the observed brightness remains 
(it actually worsens); 
so, the only solution for truncated models appears to be
to simultaneously increase \qH.
But, again, the brightness profile is not reproduced if the matter distribution
($N_e$ and $\epsilon$) is left unchanged; furthermore, $\epsilon$ cannot 
be increased, since it would increase too much the degree of ionization.
If $N_e$ is increased (by means of increasing either $N_0$ or $N_e^{min}$), 
fairly satisfactory solutions can be found, 
as far as we do not depart too much from the reference radiation-bounded models.

Summarizing, it appears that no substantial leakage of photons is affecting NGC 5461;
as a gross estimate, we can say that at most 20\% of photons escape from the region,
and that the radiation-bounded model is a good approximation to the NGC 5461 case.
Other solutions may still be possible, but they involve radical changes in the 
nature of the ionizing source (namely, a much softer radiation field),
and we did not explore them in this work.
And, of course, these results have been obtained for the particular case studied,
while for other \hii\ regions the situation might well be different.

\subsection{SF Law}\label{SFR}

As mentioned earlier, the observed \ew\ values are compatible
with both a young ($t\sim 3$ Myr) burst and a slightly older 
($t\sim 5$ Myr) continuous SF event.
A substantial improvement in the \sii/\oii\ ratio can be obtained with the 
CSF case, thanks to the contribution of older stars to the
formation of \sii. 
However, in the CSF scenario the \oiii/\oii\ ratio remains too high 
for all ages and metallicity values, due to the continuous replenishment
of hot, massive stars.
We consider then that the IB is a better approximation to the real SF
process going on in NGC~5461; 
an even better approximation would be obtained by adding to the IB
spectrum the low-frequency emission from an older population.

\subsection{IMF}

Since the slope of the IMF has no major effect on the results,
we chose the standard Salpeter value $1+x=2.35$.
Regarding \mup, we chose a relatively low \mup\ value, since higher values
yield far too high ionization degrees.
For instance, if in our reference model we modify \mup\ from
$80$ M$_{\odot}$ to $100$ M$_{\odot}$ or to $120$ M$_{\odot}$,
the \oiii/\oii\ ratio of the complete model changes from 0.42 to 1.37 or to 2.23 respectively;
the trend followed by the slit-biased values is even more extreme,
since, e.g., the \oiii/\oii\ ratio seen through slit B changes in the same sequence
from 1.14, to 5.64, to 13.41. 
We emphasize here that our results depend directly on the constraint enforced on 
the radial density distribution, and that many more solutions would be available
if it were dropped.
 
\section{CONCLUSIONS}\label{conclusions}

We presented a selected photoionization model for the GEHR NGC~5461.
The model is an asymmetric nebula, characterized by an off-center gaussian density
distribution, with a peak value of $N_e=$ 500 cm$^{-3}$ and different 
$\epsilon$ values in the two components.
The ionizing source is a young (3.1 Myr) burst with a Salpeter's IMF 
and \mup\ = 80 M${_\odot}$, 
containing 4000 O stars approximately
corresponding to 3000 `equivalent O7 V' stars, with the definition by Vacca (1994).
The reasons underlying our choice of the IBSF
reside mainly in the relatively low ionization degree of the nebula; 
CSF models are continuously replenished of hot, massive stars
and maintain a high ionization degree.
The age has been set taking into account the same constraint,
and it is fully consistent with the observed \ew\ value.
We estimate a total stellar mass of about
$M_*^{\rm tot} = 3 \times 10^{6}$ M$_\odot$ in the $(0.01-80)$ M$_\odot$ range,
and an ionized-gas mass lower limit of $M_{\rm gas}^{\rm tot} = 1.6 \times 10^{6}$ M$_\odot$.
Accounting for the gas in neutral and molecular form,
we find that the star-formation efficiency lies in the $3-5\%$ range.

Our results are pretty robust with respect to
reasonable variations in the input ingredients.
In particular, we are confident in our estimates of the stellar population's
parameter, 
leaving open for the moment the questions of metallicity and W-R population.
We are also confident that the overall matter distribution chosen
(roughly, the amount of ionized gas found at each radius,
resulting from the interplay of density and filling factor)
is a good approximation to the real one,
while we could not assess whether the constant-filling-factor configuration
can be replaced by other gas distributions with different combinations of
density and filling factor.
We consider that the region can be satisfactorily described as
radiation bounded, and that no substantial leakage of photons is taking place. 
Stated synthetically, this comes as a consequence of the relatively low
degree of ionization, coupled with the density and brightness profiles.

We successfully reproduced the following constraints:
{\sc [O~iii]} \OIII, \ctst/\cs, and {\sc [O~ii]} \OII\ as a function of the slit aperture;
the {\sc [S~ii]} \sIIratio\ profile along the nebula;
the H$\alpha$  profile along the nebula;
the equivalent width of H$\beta$ as a function of the slit aperture;
the \rdt\ parameter.
We failed to reproduce the observed He{\sc ~ii} \csos, {\sc [O~iii]} \ctst, {\sc [O~i]} \stcc,
and {\sc [S~ii]} \ssv\ line intensities.
We attribute the failure in reproducing the {\sc [S~ii]}  \ssv\ line to an inadequate
representation of the older, cooler part of the stellar population. 
This explanation could apply also to the \csos\ case: 
however, the low W-R statistics, and the theoretical problems still existing
in the modeling of the atmosphere of these stars, might provide alternative explanations.

Our study also demonstrates the following points:
\begin{enumerate}
 \item Photoionization modeling can be used to constrain the properties of {\sc H~ii} regions,
       as well as to assess our understanding of physical processes in ionized plasma.
 \item A detailed modeling should take into account as many constraints as possible,
       and the observed properties should be reproduced not only globally, but as locally as possible.
 \item Resolved spectroscopic studies are needed to provide modelers with large, homogeneous, and exhaustive
       data sets for each modeled region.
\end{enumerate}

\acknowledgements{The authors acknowledge Grazyna Stasi\'nska for many challenging comments, 
which greatly contributed to improve this work.
V.L. would like to thank Daniel Schaerer and Corinne Charbonnel for extensive support,
and the Observatoire Midi-Pyr\'en\'ees de Toulouse for providing facilities during this research.
V.L. is also grateful to Miguel Cervi\~no for many excellent suggestions.}

\clearpage

\begin{figure}
\plotone{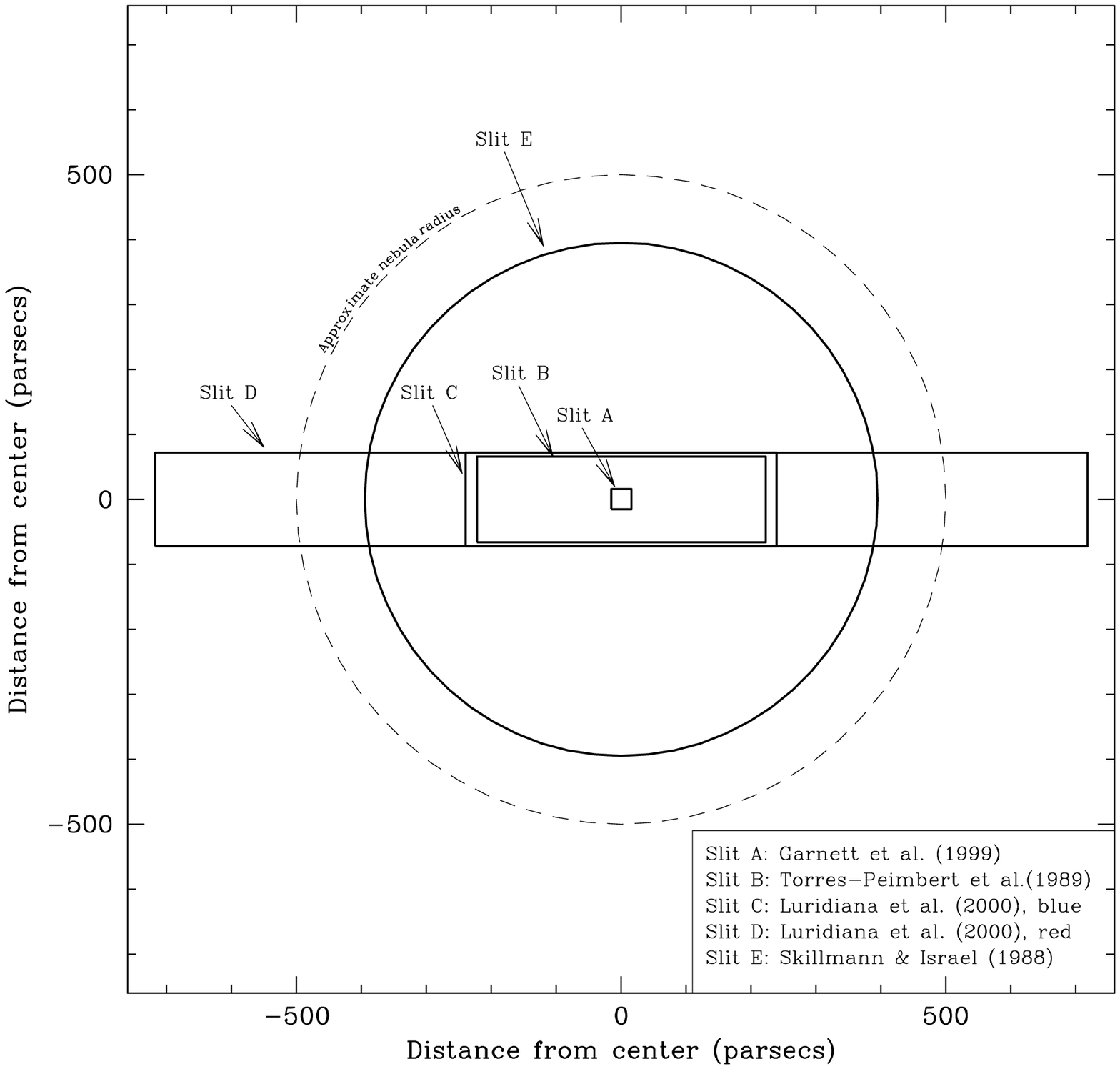}
\end{figure}

\begin{figure}
\plotone{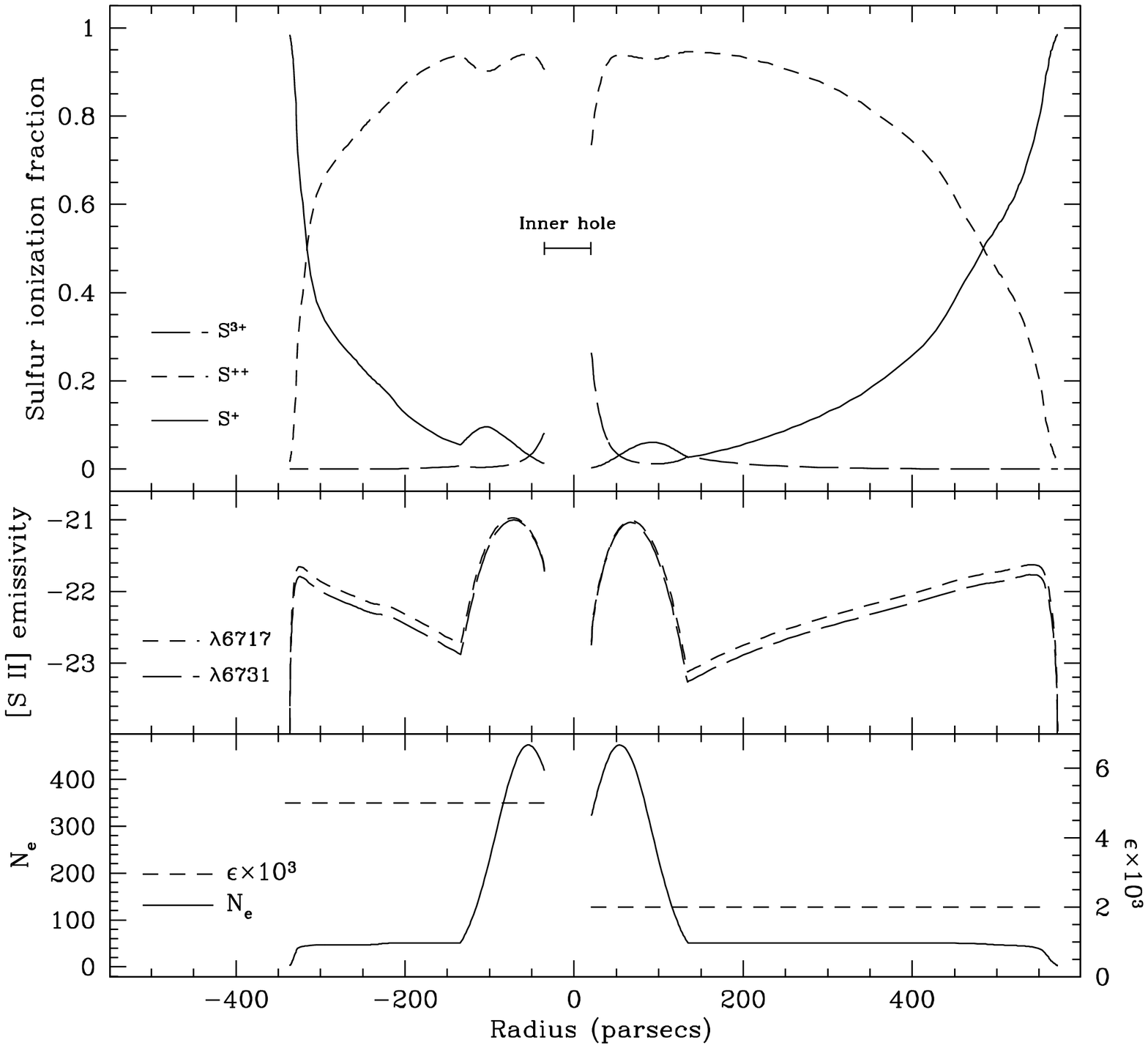}
\end{figure}

\begin{figure}
\plotone{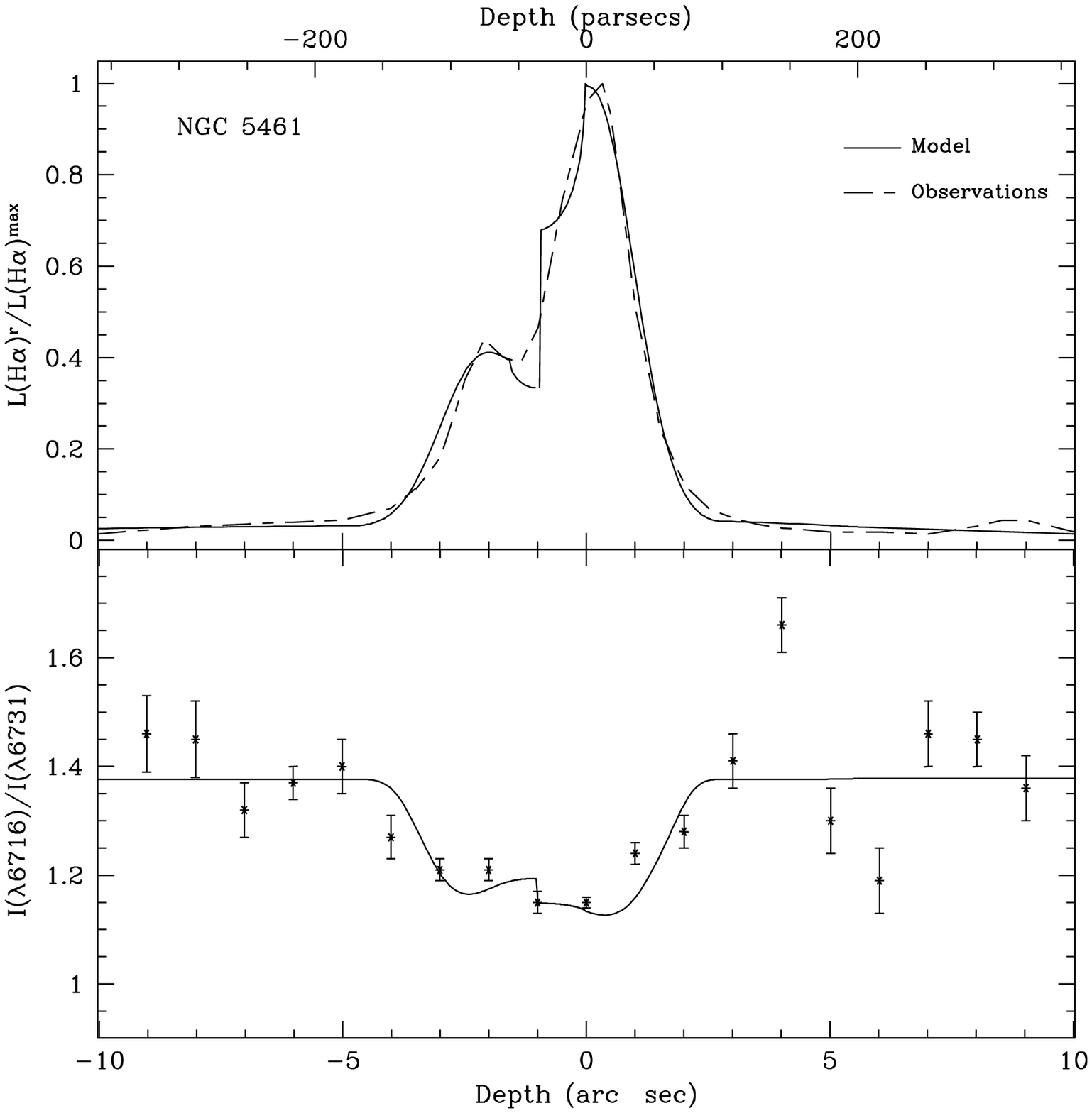}
\end{figure}

\figcaption[f1.eps]{Density distribution of a typical model. $N_{\rm 0}=500$ cm$^{-3}$, $\delta=54$ pc, $r_{\rm 0}=54$ pc. 
$\epsilon$ is $0.002$ on the left side, $0.005$ on the right.
The size and assumed position of the slits used in the subsequent analysis are also shown.\label{f_models}}

\figcaption[f2.eps]{Selected properties of our best-fit model.
Upper panel: Ionization fractions of S$^{+}$, S$^{++}$, and S$^{3+}$.
Middle panel: [{\sc S~ii}] $\lambda\lambda 6717, 6731$ emissivities as a function of radius;
note how the density profile enhances the [{\sc S~ii}] contribution near the center,
where the [{\sc S~ii}] ionization fraction is low.
Lower panel: Density and filling factor as a function of radius.\label{f_structure}}

\figcaption[f3.eps]{Upper panel: H$\alpha$ brightness profile of the best-fit model (solid line), compared to the observed
profile by Casta\~neda et al. (1992) (dot-dashed line). Lower panel: \sIIratio\ ratio profile of the best-fit model,
superposed to the observational data by Casta\~neda et al. (1992) (asterisks). The zero-point of the computed profiles has been shifted
to $r=-1''$ to make it coincide with the zero-point of the observational data.\label{f_profile}}

\end{document}